\documentclass[twocolumn,aip]{revtex4}
\usepackage{dcolumn}
\usepackage{bm}
\usepackage{graphicx}

\newcommand{ \be}{\begin{equation}}
\newcommand{ \ee}{\end{equation}}
\newcommand{\beq}{\begin{eqnarray}}
\newcommand{\eeq}{\end{eqnarray}}
\newcommand{\bem}{\begin{pmatrix}}
\newcommand{\eem}{\end{pmatrix}}
\newcommand{\bmx}{\begin{array}}
\newcommand{\emx}{\end{array}}

\begin{document}

\title{Three dimensional complex plasma structures in a combined radio frequency and direct current discharge}
\author{S. Mitic,$^1$ B. A. Klumov,$^2$ S. A. Khrapak,$^{1,2}$ and G. E. Morfill$^1$}
\address{
$^1$Max-Planck-Institut f\"ur extraterrestrische Physik, D-85741 Garching, Germany\\
$^2$Joint Institute for High Temperatures, Moscow, 125412, Russia }

\date{\today}

\begin{abstract}

We report on the first detailed analysis of large three dimensional (3D) complex plasma structures in experiments performed in pure rf and combined rf+dc discharge modes. Inductively coupled plasma (ICP) is generated by an rf coil wrapped around the vertically positioned cylindrical glass tube at a pressure of 0.3 mbar. In addition, dc plasma can
be generated by applying voltage to the electrodes at the ends of the tube far from the rf coil. The injected monodisperse particles are levitated in the plasma below the coil. A scanning laser sheet and a high resolution camera are used to determine the 3D positions of about $10^5$ particles. The observed bowl-shaped particle clouds reveal coexistence of various structures, including well-distinguished solid-like, less ordered liquid-like, and pronounced string-like phases. New criteria to identify string-like structures are proposed.
\end{abstract}


\maketitle

\section{Introduction}

Complex (dusty) plasmas -- systems consisting of highly charged micron-size particles in a neutralizing plasma background -- exhibit an extremely rich variety of interesting phenomena \cite{Book, MorfillModPhys}, including phase transitions \cite{Thomas1994, Chu1994, Hayashi1994,  Melzer1994, ufn}, solid-liquid interfaces and crystallization front propagations \cite{Milenko}, fluid-like instabilities \cite{fluid}, noctilucent clouds~\cite{nlc}, etc. Amongst these formation and evolution  of three-dimensional well-ordered structures are of particular interest~\cite{3da,3db,3dc,Mitic,ppcf,epl,ufn,CrystPRL,CrystPRE}.
This is mainly because high temporal and spatial resolution allows us to investigate these structures along with various related phenomena at the individual particle level. Fully resolved quasi-atomistic (undamped) particle dynamics provides new insight into natural atomic and molecular systems, whose dynamics cannot be resolved
in such detail.

Since the discovery of plasma crystals \cite{Thomas1994,Chu1994,Hayashi1994,Melzer1994}
many experiments have been performed to quantify the different states of complex plasmas by
analyzing the scattered laser-light of the microparticles that were introduced in the
plasma. Most of these experiments were related to two dimensional (2D)
systems (e.g. \cite{Thomas1994,Chu1994,Hayashi1994,Melzer1994,Milenko,knapek}). Only
a few experiments were focused on the analysis of large ($\simeq 10^5$particles) 3D dusty structures \cite{3da,3db,3dc,Mitic,ppcf,epl,ufn,CrystPRL,CrystPRE}
mostly performed in parallel plate rf discharges.


Here we report on the first fully 3D reconstruction of a large microparticle cloud 
in inductively coupled electrodeless rf (or rf+dc) plasma in cylindrical geometry and detailed analysis of 3D structural data. In contrast to other typical complex plasma experiments in dc discharge tubes (see e.g.~\cite{Lipaev}) the particles in the present experiment are not levitated in standing striations, where strong variations in the electric field (and other related plasma parameters) lead to inhomogeneities in particle clouds even on small scales. Instead, the particles are confined in a vertically mounted discharge tube, where levitation becomes possible due to an rf (or rf+dc) electric field and an upward neutral gas flow balancing the particle gravity and the ion drag force.\\

\begin{figure}[tbp]
\centering
\includegraphics [width=6cm] {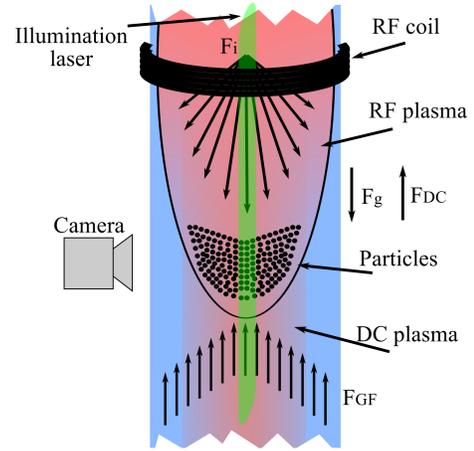}
\caption{(Color online) Sketch of the setup and of the force balance on the particles levitating in the combined rf and dc plasmas where $F_i$ is ion drag force, $F_g$ gravity, $F_{DC}$ electric force by dc plasma and $F_{GF}$ is neutral drag force (due to the gas flow).}
\label{sketch}
\end{figure}

\section{Experiment}

The experiment is performed in the ``PK-4'' setup designed as a laboratory prototype of the next generation complex plasma experimental facility onboard the ISS. The heart of this setup is a vertically oriented ``U''-shaped glass tube of 30 mm in diameter and 300 mm working length. The setup is equipped with all standard systems for operating the plasma, vacuum pumps, flow controllers, high voltage supplies for dc and rf driven plasmas, and a system for particle observation: illumination laser spread into a thin sheath and a fast camera perpendicular to it \cite{markus}. Specific for this experiment is the system for particle scanning, where illumination laser and camera are moved together by a motor for scanning through the depth of the particle cloud, described previously in \cite{Mitic}. Figure 1 shows the sketch of the experiment indicating the position of the particle cloud and the main forces acting on the particles.

\begin{figure}[t]
\centering
\includegraphics [width=8cm] {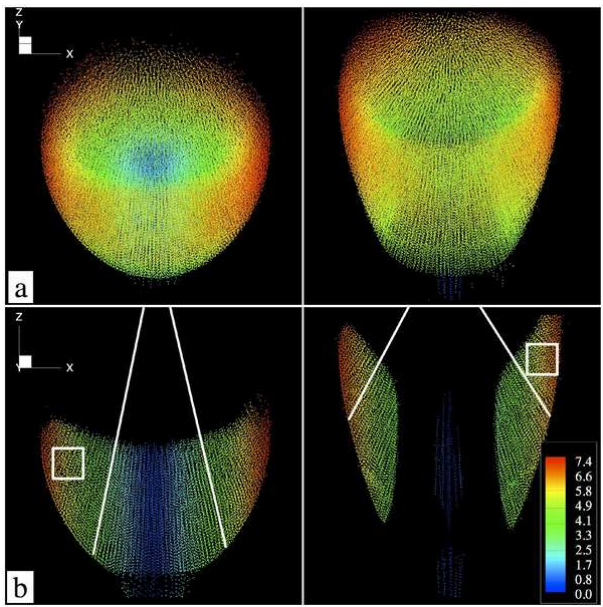}
\caption{(Color online) Dust structures observed in the pure rf (left panel) and combined rf+1mA dc (right panel) discharge with general (a) and side (b) views. Particles are color-coded by their radial distance from the tube axis (the distance is in mm units). The white squares show relatively homogenous regions used for the structural analysis. The lines indicate approximately the local orientations of the (string-like) structures. Considered structures consist of about $10^5$ particles, which is comparable to the biggest complex plasma systems analyzed so far.}
\label{boundaries}
\end{figure}

The inductively coupled plasma is sustained by an rf current at frequency of 81.36 MHz driven through the wire looped around the middle of the tube. The input power of the rf coil is 1.6 W, creating an elliptically shaped plasma. In addition, plasma can be generated through the whole tube by applying a dc voltage to the electrodes at its ends.
Experiments are performed in a mixture of argon (90\%) and oxygen (10\%) at a pressure of 0.3 mbar and an upwards gas flow of 0.25 sccm. Although similar global behavior is observed in pure argon and neon plasmas full structural analysis was done for this Ar:O gas mixture. Gas flow is used in order to avoid the plasma degradation and to provide additional compensation for the particle gravity. This additional compensation is important to position the particle cloud in the more homogeneous plasma, where the particle system is less disturbed by plasma-related instabilities.

Melamine formaldehyde microparticles of 2.55 $\mu$m in diameter are injected from the top of the tube. Due to the balance of the upward electrical and neutral drag (associated with the gas flow) forces with the downward gravity and ion drag forces, particles can levitate in a certain region below the rf coil (see Figs.~\ref{sketch} and \ref{boundaries}).

In the basic experiment particles are levitated in pure rf plasma with the mentioned constant gas flow. Approximately a minute after the particle injection a scan through the particle cloud is performed with 26 frames per mm (and 33 frames per second). After the scan a dc voltage is applied to the electrodes, creating a current limited plasma. The dc electric field is oriented to additionally compensate the particle weight. The scan through the particle cloud is repeated for the new plasma conditions. Current is changed to 0.2, 0.3, 0.4, 0.6, 1 mA and particles positions are determined for each current. The third dimension (perpendicular to the field of view) of particles positions are reconstructed based on the intensity profile of the illumination laser as described in \cite{Mitic}. During scans every particle position is reconstructed in 7 to 10 frames from which we determined the average particle position in all three dimensions. The average vibrational motion of particles during its detection is in most cases limited to one or two pixels, corresponding to 0.02 to 0.04 mm which is much smaller compared to the average interpaticle distance of around 0.2 mm. These numbers for particle displacements are in good agreement with our estimates of particle diffusion for parameters relevant to the present experiment. The Epstein damping rate~\cite{Epstein}, characterizing frictional dissipation in grain-neutral collisions is $\nu_{\rm fr}\simeq 130$ s$^{-1}$. The dimensionless parameter, which can appropriately characterize the damping strength is the so called ``damping index'' \cite{PoP2012}, $\xi = \nu_{\rm fr}\Delta\sqrt{M/T}$, where $\Delta$ is the interparticle distance, $M$ is the particle mass, and $T$ is the kinetic temperature of the particle system [in the following we assume that $T$ is equal to the neutral gas (room) temperature]. Physically, $\xi$ is the ratio between the characteristic interparticle spacing and mean ballistic free path of the particle. The particle motion is merely ballistic (Newtonian) for $\xi\ll 1$, while for $\xi\gg 1$ it is overdamped (Brownian). For our parameters we get $\xi\simeq 40$, which indicates that the particle motion approaches overdamped regime. The characteristic value of the diffusion coefficient for the system of particles containing solid and fluid phases can be roughly taken as the diffusion coefficient at the fluid-solid phase transition, $D\simeq 0.1 D_0$, where $D_0=T/(M\nu_{\rm fr})$ is the bare Brownian diffusion coefficient (for non-interacting particles). This relation is known as the dynamical criterion for freezing \cite{Lowen}. In our conditions $D_0\sim 3\times 10^{-5}$ cm$^2$/s and hence $D\sim 3\times 10^{-6}$  cm$^2$/s. The characteristic displacement during $\tau\sim 0.3$ s (10 frames) is $\ell\sim\sqrt{6D\tau}\sim 0.02$ mm, in good agreement with the observations. The characteristic time required to travel one interparticle distance, $\tau_{\Delta}\sim \Delta^2/6D\sim 25$ s, is longer than the scanning time.


\begin{figure}[tbp]
 \centering
 \includegraphics [width=8cm] {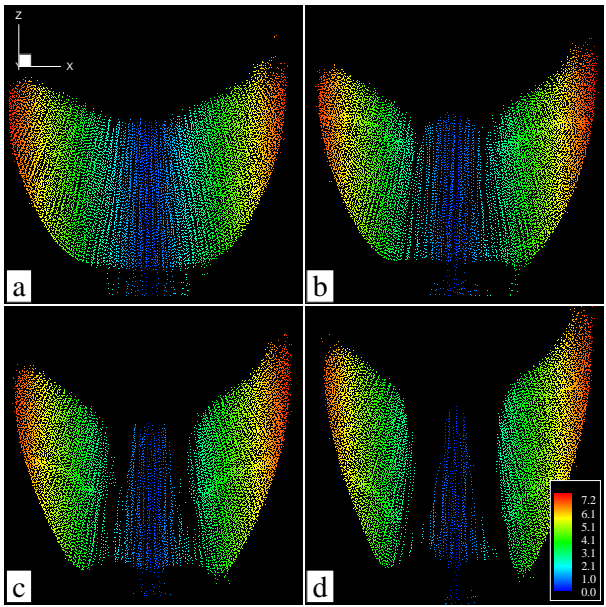}
 \caption{(Color online) Axial crossections of the particle cloud levitated in combined rf and dc discharge with different dc currents (a) 0.1 mA, (b) 0.2 mA, (c) 0.4 mA and (d) 0.6 mA. It is clearly visible that the global distribution of the particles is systematically effected by the dc current increase, slowly decreasing the number of particles levitating along the center of the tube and gradually forming the void around them. Particles are color coded by the radial position indicated by the color (distances are in mm units).}
 \label{4fig}
\end{figure}

Dust structures observed in pure rf plasma have a bowl-like shape. With the applied dc voltage the particle density near the tube center decreases, and a void is gradually formed when the dc current increases (see Figure ~\ref{boundaries} and~\ref{4fig}). We propose the following explanation for this observation. The application of dc voltage shifts the maximum of the rf plasma production down with respect to the position of the rf coil (i.e. in the direction of the dc electric field). This is evidenced by the analysis of the distribution of the integral plasma glow along the tube axis, obtained using a video imaging. This implies that the plasma density increases below the coil (and decreases above it) in comparison to the pure rf generation case. The result is that the relative importance of the ion drag force in this region increases. Increasing the absolute magnitude of the ratio of the ion drag force to the electric force (which should not necessarily exceed unity) has a consequence that the particle levitation becomes impossible in some vicinity around the tube axis.
Very close to the tube axis some particles can still levitate due to the upward neutral gas flow, which has a parabolic profile and produces maximal effect on the axis. Levitation remains also possible sufficiently far from the axis, where the plasma density decreases and the total ion flow velocity increases due to the additional radial component of the discharge electric field (both effects lower the relative importance of the ion drag force~\cite{PoP2005}).
For intermediate radial position no levitation is possible and a void is formed. This explanation is in qualitative agreement with observations, cf. Fig.~\ref{boundaries} and~\ref{4fig}.

\section{Structural properties}

Irrespective of the applied dc voltage, the observed clouds of particles are not very homogeneous. Even in a pure rf regime, typical interparticle separations in the central part of the cloud can exceed those in the peripheral regions close to the cloud boundaries by a factor of about two. For this reason, a relatively small peripheral part of the cloud sketched in Fig.~\ref{boundaries}(b) has been chosen for the detailed analysis of the particle systems structural properties. This part is sufficiently small (especially its radial extent) so that the system inside is reasonably homogeneous. At the same time, it contains enough particles ($\simeq 10^4$) to yield reasonable statistics. For the rest of the article we compare the structural arrangement of the levitated particle cloud under two extreme conditions, pure rf plasma and combined rf and dc plasma with maximal dc current of 1mA.

\begin{figure}[tbp]
\centering
\includegraphics [width=8cm] {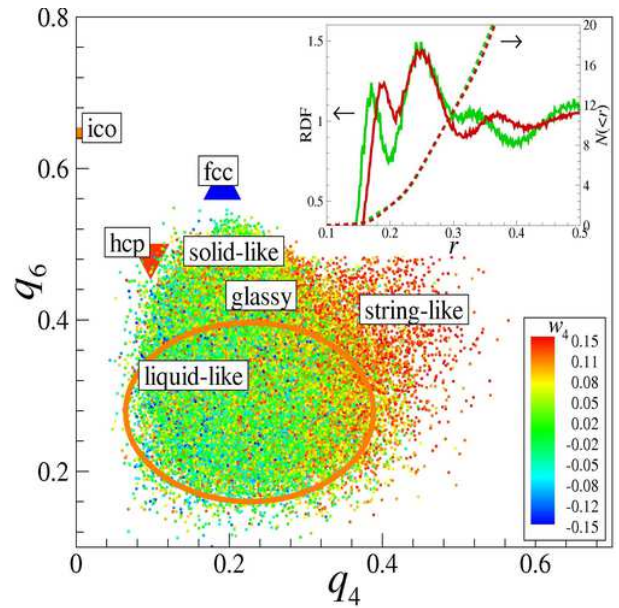}
\caption{(Color online) The typical particle distribution on the plane of bond order parameters $q_4$ and $q_6$
 (for pure rf plasma). Particles are color-coded by the $w_4$ value. Rotational invariants $q_4,~ q_6,~w_4$ were calculated by using 12 nearest neighbors; those for perfect fcc/hcp/ico lattice types are also indicated. The distribution reveals presence of
 solid-like (hcp-like and fcc-like), weakly disordered (glassy), liquid-like and string-like phases in the cloud.
 Inset shows the radial distribution function (RDF) for rf (solid red line) and rf+1mA dc (solid green line) plasma. The cumulative distributions $N(<r)$ (mean number of particles located inside sphere of radius $r$ in mm units) of the RDFs are shown by dashed lines. Both the cumulative distribution $N(<r)$ and splitting of the first maximum of the RDF indicate the presence of string-like structures.}
\label{lo}
\end{figure}

To determine the local structural properties of the three-dimensional particle system we use the bond order parameter method \cite{SteinPRL,SteinPRB}, which has been widely used to characterize order in simple fluids, solids and glasses, hard-sphere systems, colloidal suspensions, 3D complex plasmas (see e.g. Refs.~\cite{CrystPRL,CrystPRE,KlumovPRB} and references therein). In this method, the rotational invariants of rank $l$ of both second $q_l(i)$ and third $w_l(i)$ order are calculated for each particle $i$ in the system from the vectors (bonds) connecting its center with the centers of the $N_{\rm nn}(i)$ nearest neighboring particles:
\be
q_l(i) = \left ( \frac{4 \pi}{(2l+1)} \sum_{m=-l}^{m=l} \vert~q_{lm}(i)\vert^{2}\right )^{1/2},
\ee
\be
w_l(i) = \hspace{-0.8cm} \sum\limits_{\bmx {cc} _{m_1,m_2,m_3} \\_{ m_1+m_2+m_3=0} \emx} \hspace{-0.8cm} \left [ \bmx {ccc} l&l&l \\
m_1&m_2&m_3 \emx \right] q_{lm_1}(i) q_{lm_2}(i) q_{lm_3}(i),
\label{wig}
\ee
\noindent
where $q_{lm}(i) = N_{\rm nn}(i)^{-1} \sum_{j=1}^{N_{\rm nn}(i)} Y_{lm}({\bf r}_{ij} )$, $Y_{lm}$ are the spherical harmonics and ${\bf r}_{ij} = {\bf r}_i - {\bf r}_j$ are vectors connecting centers of particles $i$ and $j$. In Eq.(\ref{wig}) $\left [ \bmx {ccc} l&l&l \\ m_1&m_2&m_3 \emx \right ]$ denote the Wigner 3$j$-symbols, and the summation in the latter expression is performed over all the indexes $m_i =-l,...,l$ satisfying the condition $m_1+m_2+m_3=0$. The calculated rotational invariants  $q_i,~w_i$ are then compared with those for ideal lattices \cite{SteinPRL, SteinPRB, ufn}. Here, we are specifically interested in identifying face-centered cubic (fcc), hexagonal close-packed (hcp), icosahedral (ico) and body-centered cubic (bcc) lattice types and, therefore, use the invariants $q_4$, $q_6$, $w_4$, $w_6$ calculated using the {\it fixed numbers} of $N_{\rm nn}=12$ (fcc/hcp/ico) and $N_{\rm nn}=8$  (bcc) nearest neighbors, respectively. A particle whose coordinates in the 4-dimensional space $(q_4,q_6,w_4,w_6)$ are sufficiently close to those of the ideal fcc (hcp, ico, bcc) lattice is counted as fcc-like (hcp, ico, bcc-like) particle.

An example from this analysis is presented in Fig.~\ref{lo}, which shows representative particle distributions on the plane ($q_4$, $q_6$) taken for the particle cloud observed in pure rf plasma. The distribution of particles versus rotational invariants $q_4$, $q_6$ reveals the presence of solid-like (mostly hcp-like), string-like (big values of $q_4$ and $w_4$), medium ordered glassy ($0.4 \le q_6 \le 0.44$) and weakly ordered liquid structures ($q_4 \le 0.4$ and $q_6 \le 0.4$ ) in the analyzed part of the cloud. The corresponding domains are indicated in the plot.

\begin{figure}[tbp]
\centering
\includegraphics [width=8cm] {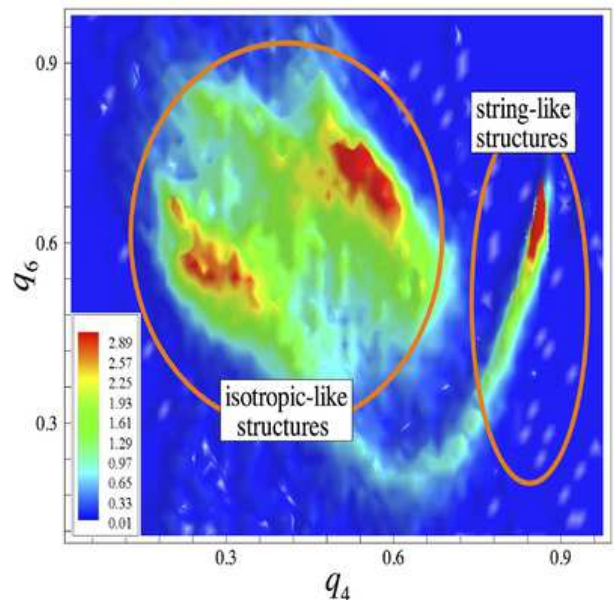}
\caption{(Color online) Typical particle distribution on the plane of the order parameters $q_4\--q_6$ (calculated by using 6 nearest neighbors). The domains occupied by both the quasi-isotropic and string-like structures (consisting of $N_{\rm SLS}> 7$ microparticles) are indicated. The color coding from blue to red corresponds to the increasing probability of finding a particle in a given region of ($q_4$, $q_6$) plane.}
\label{anis}
\end{figure}

\begin{figure}[tbp]
\centering
\includegraphics [width=8cm] {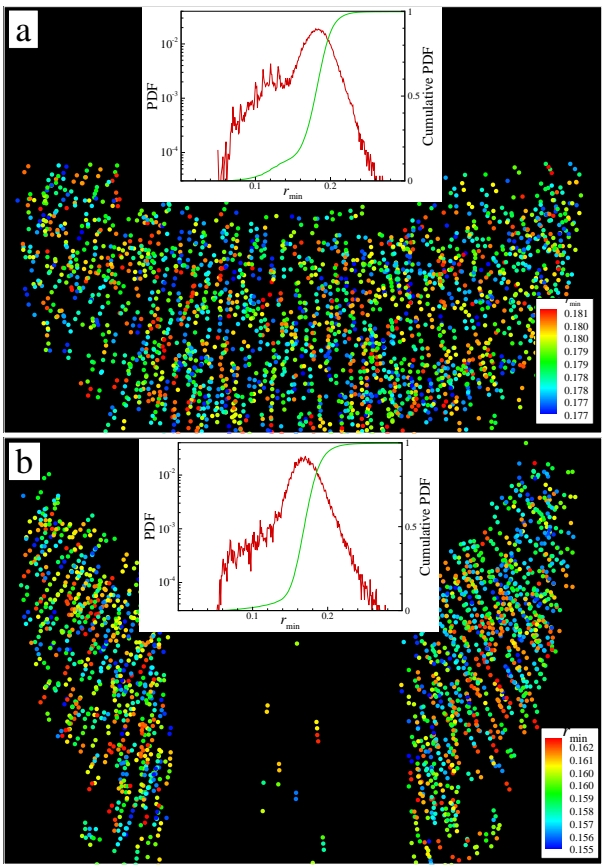}
\caption{(Color online) Observed string-like structures in pure rf (a) and rf+1mA dc (b) plasma.
Particles are color-coded  by the parameter $r_{\rm min}$(mm) which is the distance to the nearest neighbor.
Inset shows corresponding probability distribution functions. The cumulative distribution (green line) are also plotted. The PDF reveal shoulder-like behavior at distances $r$ far less than the averaged interparticle distance, indicating the presence of string-like structures in the system.}
\label{sls}
\end{figure}

The inset in Fig.~\ref{lo} shows the radial distribution function (RDF) $g(r)$ for both pure rf (solid red line) and combined rf+1mA dc (solid green line) plasma. Cumulative distributions $N(<r)$ (mean number of neighboring particles in the sphere of radius $r$) for these RDFs are also plotted (dashed lines). Both $g(r)$ and $N(<r)$ indicate the presence of the string-like structures (SLSs) in the particle cloud: The splitting of the first RDF maximum (at $N \simeq 12$) is one of the indicators of developed SLSs in the system (see \cite{ppcf}).

In view of the ubiquitous character of string-like structures in various complex plasma experiments (e.g.~\cite{Book,Arp}), including the present one, it would be of value to develop simple and convincing indicators that can be used to identify the transition from isotropic to anisotropic (SL) structures. Here we put forward and briefly discuss three related possibilities.

First, we consider a new descriptor of the SL structures, which is related to the distribution of particles in the plane ($q_4$, $q_6$), where the rotational invariants $q_4$, $q_6$ are calculated using a fixed number of 6 nearest neighbors (instead of 12 and 8 to identify isotropic crystalline structures considered above).
This choice of the number of nearest neighbors used for detection of string like structures is to some extend arbitrary. In the analyzed case, observation of 6 nearest particles yields good statistic and clear identification of ``good'' strings: In an ideal case when all 7 particles tend to lie on the same line, the value of $q_4$ tends to unity.
Using smaller number of nearest neighbors results in higher values of $q_4$ and $q_6$. Nevertheless, the pronounced peak in the $q_4$ values, well separated from the values characterizing the rest of the system would give clear identification of the anisotropy. A typical example of such distribution (for the regime of pure rf plasma) is shown in Fig.~\ref{anis}.  Clearly, the existence of SL structures (consisting of $N_{\rm SLS} \gtrsim  7$ microparticles) can be easily identified.

Another useful measure of the system anisotropy is the distribution function of the distance $r_{\rm min}$ between a test particle and its nearest neighbor. The corresponding probability distribution functions (PDFs) are shown in Fig.~\ref{sls} together with the overview of the particle clouds. The presence of the SL structures is evidenced by a non-gaussian behavior of these PDFs. Thus, the shape of the PDFs of $r_{\rm min}$ turns out to be a useful tool to identify system anisotropy, in addition to the widely used scaling-index method~\cite{asi, ppcf}.

\begin{figure}[tbp]
\centering
\includegraphics [width=8cm] {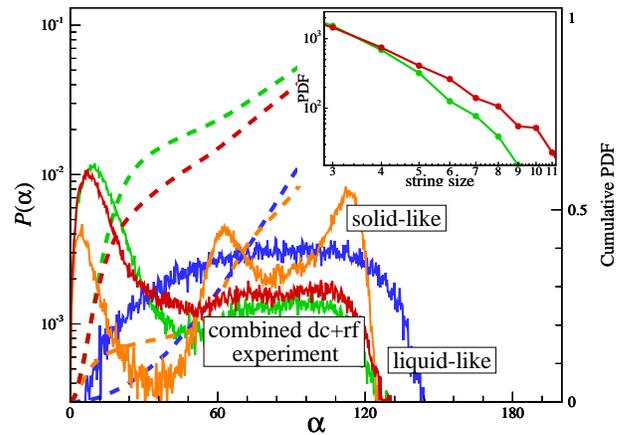}
\caption{(Color online) Probability distribution function $P(\alpha)$ in an rf (solid red line) and rf+ rf+1mA dc (solid green line) plasmas.
Similar PDFs are also plotted for isotropic liquid-like (the data set is taken from \cite{ufn}, solid blue line) and solid-like (the data set is taken from \cite{ppcf}, solid orange line) 3D complex plasmas. Corresponding cumulative distributions are also plotted by dashed lines of the same color.
The inset shows the distribution of the string-like structures over the number of particles they contain for rf (solid red line) and rf+1mA dc (solid green line) complex plasmas.}
\label{slssp}
\end{figure}

An important characteristic of SL structures is the distribution over the number of particles they contain. To find this distribution we define two nearest neighbors for each particle $i$, their positions are characterized by ${\bf r}_{i,1}$, ${\bf r}_{i,2}$ and ${\bf r}_i$, respectively. We then calculate the angle $\alpha$
between the vectors ${\bf r}_i - {\bf r}_{i,1}$ and ${\bf r}_{i,2} - {\bf r}_i$. If $\alpha$ is sufficiently small ($\alpha \le \alpha^*$), it is postulated that the particles belong to the same string (in practice $\alpha^*\approx 10^{\circ}$ is used, the sensitivity to the exact value of $\alpha^*$ is rather weak). Clustering algorithm (see~\cite{ufn}) is used to define the total number of particles in the string and to obtain the related distribution of SL structures over the number of particles. Figure~\ref{slssp} shows typical probability distributions $P(\alpha)$ for the observed complex plasmas for rf (red line) and rf+1mA dc (green line) plasma. Additionally, we plotted $P(\alpha)$ for the solid-like and isotropic liquid-like complex plasma systems (calculated from the data of Refs.~\cite{ufn} and~\cite{ppcf}) to emphasize structural anisotropy observed in the present experiments. The inset in Fig.~\ref{slssp} shows distributions of the observed SLSs over the number of particles they contain. These distributions clearly reveal a power law-like behavior, indicating the absence of any spatial scales characterizing the anisotropy of the particle systems under consideration.

\section{Conclusion}

To conclude, we present first detailed analysis of large 3D complex plasma structures in rf and combined rf+dc generated plasmas. Structural analysis reveals a rather complicated mixture of different phases in the observed complex plasmas: presence of solid-like, liquid-like and string-like structures in the system is well established. Variations of the dc current significantly change the global structure of the particle cloud, but do not affect the local structural properties significantly. Finally, we propose new indicators to identify and quantify string-like structures appearing in complex plasmas. One of the observations from the present experiment is that the identified SLSs exhibit scale-free power-law-like distributions over the number of particles they contain.

\begin{acknowledgements}
This work was supported by DLR under Grants 50WM0804 and 50WM1150 and by ESA within the frame of PK-4 project.
This study was also partly supported by the Presidium and Division of Physical Sciences of the
Russian Academy of Sciences; the Ministry of Education and Science of the Russian Federation; and the Russian
Foundation for Basic Research, Projects no. 13-02-00-913 and 13-02-01099.
\end{acknowledgements}

\end{document}